\documentclass[%
pra, superscriptaddress, twocolumn,
amsmath, amssymb, aps, pra, 
floatfix, ]{revtex4-1}
\usepackage{nicefrac}
\usepackage{graphicx}
\usepackage{bm}
\usepackage{todonotes}
\usepackage[protrusion=true,expansion=true,tracking=true]{microtype}
\usepackage{hyperref}
\usepackage[capitalize]{cleveref}
\usepackage{dsfont}
\usepackage{braket}
\usepackage{xcolor}
\usepackage{cleveref}
\usepackage[caption=false]{subfig}
\usepackage{notes2bib}
\bibnotesetup{
note-name = ,
use-sort-key = false
}

\crefname{section}{Sec.}{Sec.}

\newcommand{\mycaption}[2]{\caption[#1]{\emph{#1} #2}}

\begin{document}
\title{Splitting of topological charge pumping in an interacting two-component fermionic Rice-Mele Hubbard model}

\author{E. Bertok}
\affiliation{Institute for Theoretical Physics, Georg-August-Universit\"at G\"ottingen,
Friedrich-Hund-Platz 1, 37077 G\"ottingen, Germany}


\author{F. Heidrich-Meisner}
\affiliation{Institute for Theoretical Physics, Georg-August-Universit\"at G\"ottingen,
Friedrich-Hund-Platz 1, 37077 G\"ottingen, Germany}

\author{A. A. Aligia}
\affiliation{Institute for Theoretical Physics, Georg-August-Universit\"at G\"ottingen,
Friedrich-Hund-Platz 1, 37077 G\"ottingen, Germany}
\affiliation{Instituto de Nanociencia y Nanotecnolog\'{\i}a CNEA-CONICET,
Centro At\'{o}mico Bariloche and Instituto Balseiro, 8400 Bariloche, Argentina}

\date{\today}
\begin{abstract}
A Thouless pump transports an integer amount of charge when pumping adiabatically around a singularity.
We study the splitting of such a critical point into two separate critical points by adding a Hubbard interaction. Furthermore, we consider extensions to a spinful Rice-Mele model, namely a staggered magnetic field or an Ising-type spin coupling, further reducing the spin symmetry.
The resulting models additionally allow for the transport of a single charge in a two-component system of spinful fermions, whereas in the absence of interactions, zero or two charges are pumped. In the SU(2)-symmetric case, the ionic Hubbard model is visited once along pump cycles that enclose a single singularity. Adding a staggered magnetic field additionally transports an integer amount of spin while the Ising term realizes a pure charge pump.
We employ real-time simulations in finite and infinite systems to calculate the adiabatic charge and spin transport, complemented by the analysis of gaps and the many-body polarization to confirm the adiabatic nature of the pump. The resulting charge pumps are expected to be measurable in finite-pumping speed experiments in ultra-cold atomic gases, for which the SU\((2)\) invariant version is the most promising path. We discuss the implications of our results for a related quantum-gas experiment by Walter \textit{et al.} [arXiv:2204.06561].

\end{abstract}

\maketitle
\section{Introduction}

The advent of ultra-cold quantum-gas experiments \cite{Fisher1989,Jaksch1998,Greiner2002a} has opened the possibility of directly probing quantum many-body systems on lattice models to a high precision.
Strongly interacting systems can give rise to many exotic phases that often arise due to the competition between different energy scales \cite{Bloch2008a}.
An open question in condensed matter theory is the precise interplay between many-body physics and topology. Thouless charge pumps \cite{Berg2011, Niu1984, Thouless1983} provide a practical framework to study interacting topological systems in a reduced spatial dimension due to their highly controllable experimental realizations.
Experimentally, Thouless pumps have been realized in ultra-cold atoms for both bosons \cite{Lohse2016} and fermions \cite{Nakajima2016, nakajima2021,Walter2022}, as well as in photonic systems \cite{Cerjan2020}.

\begin{figure}[]
  \centering
  \includegraphics{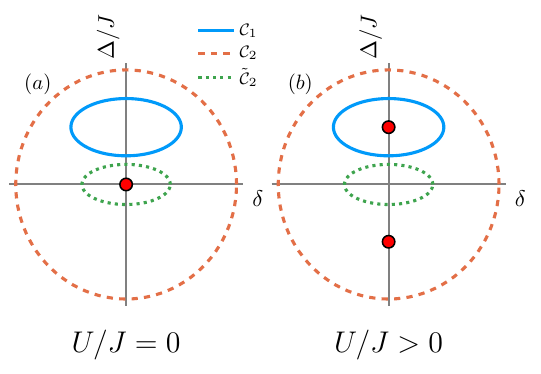}
  \mycaption{Sketch of pump cycles in the Rice-Mele-Hubbard model.}{ $\Delta$ is
  the amplitude of the staggered potential. $\delta$ is the hopping modulation.
  Two types of path are considered: $\mathcal C_1$ is centered around a point on the
  $\Delta$-axis with a nonzero offset, while $\mathcal C_2$ and $ \tilde{\mathcal C_2}$ are centered around the
  origin. (a) For $U = 0$, two charges are pumped along $\mathcal C_2$ and $\tilde{\mathcal  C_2}$ while
  no charges are pumped along $\mathcal C_1$. (b) For finite interactions, one charge
  is pumped along $\mathcal C_1$ instead. Along $\mathcal C_2$, two charges are pumped still,
  while along $ \tilde{\mathcal C_2}$, zero charges are pumped, as no singularity is encircled.}
  \label{fig:cartoon_paths}
\end{figure}

In a Thouless pump, an integer amount of charge is pumped per pump cycle when adiabatically changing parameters such that a degeneracy (or critical point) is enclosed without closing a gap.
The prototypical model for a non-interacting charge pump, the Rice-Mele model \cite{Rice1982}, has a single degeneracy at the origin as seen in \cref{fig:cartoon_paths}(a). \(\Delta\) is the strength of a staggered potential and \(\delta\) is the strength of the hopping dimerization. For a non-interacting two-component fermionic system, going once around the path \(\mathcal C_2\) pumps two particles, whereas going around \(\mathcal C_1\) pumps no particles.

Theoretically, both bosonic \cite{Hayward2018, Berg2011, Ke2017, Rossini2013, Zeng2016a, Greschner2020} and fermionic \cite{Kuno2019, Nakagawa2018, Lin2020c, Voorden2019, Esin2022} topological charge pumps have been studied.
Due to their readily available experimental realization, the interplay of Hubbard interactions and Thouless pumps in a two-component fermionic system is a key area of research. Recent works include theoretical studies of instantaneous topological measures for quantum many-body phases \cite{Stenzel2019}, the theoretical \cite{Nakagawa2018} and experimental \cite{Walter2022} study of the breakdown of topological pumping due to interactions and interaction-induced topological pumping\cite{Lin2020c,Kuno2020}. Another direction concerns the study of charge pumps in the presence of disorder \cite{Hayward2021a, Hu2020a, Ippoliti2020, Marra2020a,Wauters2019,Wang2019b,Qin2016}.

Here, we address the question whether it is possible to split the degeneracy of the non-interacting Rice-Mele model into two separate ones by adding a repulsive onsite interaction, as is sketched in \cref{fig:cartoon_paths}(b).
In this case, going along the path \(\mathcal C_1\) pumps a single charge for a finite interaction strength. With this scheme, it becomes possible to change the amount of charge pumped from zero to one by solely changing the Hubbard interaction strength. Keeping an origin-centered pumping path instead, encircling no singularities at sufficiently large \(U\) (\(\tilde{\mathcal C}_2\) in \cref{fig:cartoon_paths}\;(b)), leads to a topologically protected pumped charge of zero.

We show that this splitting is possible in three distinct situations that differ by their symmetries: (a) an SU\((2)\) symmetric fermionic model with Hubbard interactions, which can be viewed as an ionic Hubbard model (IHM) with additional alternating hopping amplitudes; (b) a model with an easy-axis spin symmetry, and (c) a model with broken \(Z_2\) symmetry. The three cases vary in the degree of adiabaticity that manifests itself in the nature of the gaps along the zero-dimerization line connecting the two critical points: The SU\((2)\) symmetric model has a vanishing many-body gap on this line, as the Mott phase of the IHM is realized between the critical points  ($\delta=0$ and $-\Delta_s \leq \Delta  \leq \Delta_s$,
where  $\pm \Delta_s$ are the values of $\Delta$ at the spin transitions discussed in \cref{sec:results} c) \cite{Manmana2004, Torio2001}. In the easy-axis case, a two-fold degenerate ground-state is obtained along this line (and in a finite region around \(\delta=\Delta=0\) in the thermodynamic limit) that is well separated from the rest of the spectrum. Finally, the \(Z_2\)-broken case has a non-degenerate ground-state separated by a robust gap from the rest of the spectrum everywhere apart from the critical points.  
We realize these three cases via (a) a Rice-Mele Hubbard model, which we take as a base model, (b) an additional Ising-type term, and (c) via an additional staggered magnetic field added to the Rice-Mele Hubbard model, respectively. This is realized by three Hamiltonians, \(\hat H_\mathrm{IH}\), \(\hat H_\mathrm{IHZ}\) and \(\hat H_\mathrm{IHB}\), respectively.

Conceptually, the \(Z_2\) broken Hamiltonian \(\hat H_\mathrm{IHB}\) has the most robust topology at the price of introducing a non-zero quantized spin current. The dimerized ionic Hubbard model \(\hat H_\mathrm{IH}\) does not feature a strict topological protection since along the pump cycle \(\mathcal{C}_1\), points exist with a vanishing spin gap. The charge gap remains open in all three cases.

As our main result, we show that we can achieve integer-quantized charge pumping around a single degenerate point. We demonstrate, via finite-time calculations, that \(\hat H_\mathrm{IHB}\) and \(\hat H_\mathrm{IHZ}\) allow for robust quantized charge pumping of a single charge per pump cycle around a single critical point.
For \(\hat H_\mathrm{IH}\), we pump through a many-body gapless phase. Still, for appropriately chosen pump cycles, the pumped charge is practically quantized on the accessible time scales, which is confirmed via time-dependent infinite-system matrix-product state methods \cite{Haegeman2016a, Zauner-stauber2018}.
We study the topology of the three models via instantaneous measures such as the energy gaps and the charge (spin-) Berry phase calculated via the many-body charge (spin-) polarization \cite{Resta1998, Aligia1999, Aligia2000}. While all models show a well-defined and smooth charge-Berry phase, the spin-Berry phase of \(\hat H_\mathrm{IHZ}\) and \(\hat H_\mathrm{IH}\) depicts a jump at a finite hopping modulation.

Our results have consequences for experiments on interacting charge pumps \cite{Walter2022}. Especially the SU\((2)\)-symmetric case is particularly simple to realize in ultra-cold-atomic gas experiments by adding a time-dependent hopping modulation to an IHM \cite{Loida2017}. Even though the pumping happens through a gapless phase in this case, we expect that integer charge pumping can be observed as we do in real-time simulations, due to finite system sizes and the resulting finite-size gaps, at least for a sequence of initial pump cycles. Our results shed additional light on the interpretation of the recent experiment by Walter \textit{et al.} \cite{Walter2022}, where the authors interpret the behavior along a cycle similar to \(\tilde{\mathcal C}_2\) as a breakdown of quantized particle pumping as a function of \(U\). We here reinterpret their results in terms of \cref{fig:cartoon_paths}\;(b) as a consequence of the singularities moving out of the cycle \(\tilde{\mathcal C}_2\) as \(U\) increases.

The paper is structured as follows. In \cref{sec:model}, we start by introducing the three models and define the relevant many-body gaps. Section \ref{sec:methods}, showcases the instantaneous and time-dependent measures and our numerical methods. In \cref{sec:results} we present our results by starting with time-dependent simulations for the pumped charge, subsequently discussing energy gaps and concluding with the Berry phases. We conclude in \cref{sec:summary} with a summary and discuss implications for a recent experimental \cite{Walter2022} and a related theoretical \cite{Nakagawa2018} study on the breakdown of topological pumping in interacting systems.

\section{Model}
\label{sec:model}

We consider a class of models of correlated fermions with a staggered potential \(\Delta\), hopping dimerization \(\delta\), a staggered magnetic field of strength \(b\) and an Ising-type term of strength \(J_z\):
\begin{align}
  \hat H = \hat H_\mathrm{IH}(\delta,\Delta) + \hat H_\mathrm{B} + \hat H_\mathrm{Z},
\end{align}
where
\begin{align}\begin{split}\hat H_\mathrm{IH}(\delta,\Delta) = &-J \sum\limits_{j=1}^{L}\sum\limits_{\alpha 
  = {\uparrow,\downarrow}}\left(1+\delta\;(-1)^j\right)\left(\hat c_{j,\alpha}^\dagger \hat c_{j+1,\alpha} 
  + \mathrm{h.c.}\right)\\
  &+\Delta\sum\limits_{j=1}^{L}\sum\limits_{\alpha 
  = {\uparrow,\downarrow}}(-1)^j \hat c_{j,\alpha}^\dagger \hat c_{j,\alpha}\\
  &+U\sum\limits_{j=1}^L\hat c_{j,\uparrow}^\dagger\hat c_{j,\uparrow}c_{j,\downarrow}^\dagger\hat c_{j,\downarrow}
\end{split}\end{align}
is the dimerized ionic Hubbard model,
\begin{align}
  \hat H_\mathrm{B} &= b \sum\limits_{j=1}^L (-1)^j \hat S_j^z
\end{align}
is a staggered magnetic field and
\begin{align}
  \hat H_\mathrm{Z} &= J_z \sum\limits_{j=1}^L \hat S_j^z \hat S_{j+1}^z
\end{align}
is an Ising spin coupling.
Here, \(\hat c_{j,\alpha}^\dagger\) creates a fermion of spin \(\alpha\in{\uparrow,\downarrow}\) on site \(j\).
The spin operators are given as \(\hat S_j^z = 1/2\left(\hat c_{j,\uparrow}^\dagger\hat c_{j,\uparrow} - \hat c_{j,\downarrow}^\dagger\hat c_{j,\downarrow}\right)\). \(L\) is the number of sites.
For \(b = J_z = \delta = 0\), the 
ionic Hubbard model (IHM) is recovered.

The phase diagram of the IHM has been studied in detail \cite{Manmana2004, Torio2001, Torio2006}.
The half-filled IHM hosts three phases, depending on the parameters \(\Delta\) and \(U\): 
A Mott insulating (MI) phase and a band insulating (BI) phase that are separated by a spontaneously dimerized (SDI) phase.

The IHM has been originally proposed \cite{Nagaosa1986} to describe the neutral-ionic transition in mixed-stack donor-acceptor organic crystals \cite{Torrance1981} and is also relevant for one-dimensional ferroelectric
perovskites \cite{Egami1993}. Its phase diagram has been determined accurately (minimizing finite-size 
effects) using the method of topological transitions \cite{Aligia2000}.
For this model, these transitions also coincide with those obtained with
the method of crossing of excited energy levels (MCEL) 
based on conformal field theory \cite{Nomura1994c,Nakamura1999,Nakamura2000}. 
There exists considerable theoretical evidence for
the existence of a bond-order wave (BOW)  phase between the Mott
insulating (MI) and band insulating (BI)  phases. This phase occurs naturally when starting in the MI phase for $\delta=0$ and adding a small $\delta$, because this term breaks the inversion symmetry (see App. \ref{sec:app}).  However, for $\delta=0$, this symmetry is broken spontaneously in the thermodynamic limit leading to a spontaneously dimerized insulator (SDI) separating the MI and BI phases. The SDI phase has been found first by bosonization \cite{Fabrizio1999}.
The IHM has recently been experimentally realized with ultra-cold atoms in a hexagonal lattice \cite{Messer2015, Tarruell2012}. The SDI has not directly been observed in experiments, although its direct measurement with superlattice modulation spectroscopy has recently been proposed \cite{Loida2017}.

In the following, we consider three families of Hamiltonians to split the charge critical points choosing the following sets of parameters:
\begin{itemize}
\item{\(\hat H_\mathrm{IH}(\delta,\Delta):\)\hphantom{\(\hat H_\mathrm{IHB}(\delta,\Delta):\)\(\hat H_\mathrm{IHZ}(\delta,\Delta):\)}}\hspace{-1.5cm} \((b,J_z) = (0,0)\)
\item{\(\hat H_\mathrm{IHB}(\delta,\Delta):\)\hphantom{\(\hat H_\mathrm{IH}(\delta,\Delta):\)\(\hat H_\mathrm{IHZ}(\delta,\Delta):\)}}\hspace{-1.5cm} \((b,J_z) = (0.5J,0)\)
\item{\(\hat H_\mathrm{IHZ}(\delta,\Delta):\)\hphantom{\(\hat H_\mathrm{IH}(\delta,\Delta):\)\(\hat H_\mathrm{IHB}(\delta,\Delta):\)}}\hspace{-1.5cm} \((b,J_z) = (0,2J)\).
\end{itemize}
The exact choice of parameters does not play a role as long as the topologies of the pump cycles \(\mathcal C_1\) and \(\mathcal C_2\) are preserved.

The pumping paths $\mathcal C_1$, $\mathcal C_2$ are parameterized via the pumping parameter \(\theta \in [0,2\pi)\):
\begin{align}
\label{eq:explicitpaths}
\mathcal C_1(\theta)&=(\Delta(\theta),\delta(\theta))=\left(\Delta_c + R_\Delta \sin{\theta},R_\delta \cos{\theta}\right),\nonumber\\
\mathcal C_2(\theta)&=(\Delta(\theta),\delta(\theta))=\left(R_\Delta \sin{\theta},R_\delta \cos{\theta}\right).
\end{align}

The time evolution along path $\mathcal C_1$ for all three models goes through a phase with non-zero BOW order parameter. In particular, \(\hat H_\mathrm{IH}\) passes through the MI phase of the IHM at $\delta=0$, surrounded by the BOW phase as soon as  $\delta \neq 0$. The same happens in recent theoretical \cite{Nakagawa2018}  and experimental \cite{Walter2022} work for the path \(\tilde{\mathcal{C}}_2\), for which a breakdown of the 
quantized charge transport was reported. 
Using a canonical transformation valid for small $J$ and $|\delta|$ and known results for 
a Heisenberg chain with alternating exchange, we see that at the MI-BOW transition, the spin gap opens as  $|\delta|^{2/3}$ while the change in polarization and the BOW order parameter behave as $\delta^{1/3}$ for small $|\delta|$. The details are in App. \ref{sec:app}.

The three phases in the IHM can be distinguished via the behavior of various many-body gaps. To distinguish the physics of the three models defined above, we introduce the following energy gaps.
We define the internal gap
\begin{align}
  \label{eq:en1}
  \Delta E_{\mathrm{int}}=E_{1}(N, S_z=0)-E_{0}(N, S_z=0)
\end{align}
as the first excitation energy keeping the total number of particles $N$ and the total spin projection $S_z=0$ constant.
We also define the charge gap
\begin{align}
  \Delta E_{C}=\left[E_{0}(N+2, S_z=0)+E_{0}(N-2, S_z=0)\right.\nonumber\\
  \left. - 2 E_{0}(N, S_z=0)\right] / 2,
  \label{eq:enC}
\end{align}
the spin gap
\begin{align}
  \Delta E_{S}=E_{0}(N, S_z=1)-E_{0}(N, S_z=0)
  \label{eq:enS}
\end{align}
and the second internal gap
\begin{align}
  \Delta E_{2}=E_{2}(N, S_z=0)-E_{0}(N, S_z=0).
  \label{eq:en2}
\end{align}

\section{Methods and Observables}

\label{sec:methods}
\subsection{Instantaneous measures}
For the instantaneous measures, we use the Lanczos method for a finite system with periodic boundary conditions up to \(L=12\).
The charge and spin gaps are calculated by searching for the lowest energy in the respective symmetry sectors. For the internal gap calculation, several low-lying eigenstates are computed.

The charge [spin] pumping is related to the charge- [spin-] Berry phases \cite{Torio2001}:
\begin{align} 
  \label{eq:berry}
  \gamma_{C, S}=&-\lim _{N \rightarrow \infty} \operatorname{Im}\left\{\operatorname { ln } \left[\prod_{r=0}^{N-2}\left\langle g\left(\Phi_{r}, \pm \Phi_{r}\right) \mid g\left(\Phi_{r+1}, \pm \Phi_{r+1}\right)\right\rangle\right.\right.\\ &\left.\left.\times\left\langle g\left(\Phi_{N-1}, \pm \Phi_{N-1}\right) \mid g(2 \pi, \pm 2 \pi)\right\rangle\vphantom{\prod_i^j}\right]\right\}\nonumber,
\end{align}
with  \(|g(2 \pi, \pm 2 \pi)\rangle=\exp \left[i 2 \pi / L \Sigma_{j} j\left(\hat n_{j \uparrow} \pm \hat n_{j \downarrow}\right)\right]|g(0,0)\rangle\),
where $N$ is the number of discretization steps for twisted boundary conditions, \(\Phi_{r}=2 \pi r / N\) and \(|g(\Phi, \Phi')\rangle\) is the ground state of the Hamiltonian in which the hopping $J$ for spin up (down) has been changed by a factor $\exp(i\Phi/L)$  $[\exp(i\Phi'/L)]$. Notice that the charge (spin-) Berry phase $\gamma_{C,S}$ depends on the pump parameter $\theta$ because the ground states \(|g(\Phi, \Phi')\rangle\) do.
The pumped charge (spin) after one pump cycle in the quasi-adiabatic limit is given by: 
\begin{align}
  \Delta Q_{C,S} = \frac{1}{2\pi}\int_0^{2\pi} d\theta\;\partial_\theta \gamma_{C,S}(\theta).
  \label{eq:deltaQberry}
\end{align}
The parameters \(\Delta\) and \(\delta\) depend on a geometrical variable \(\theta\) [see \cref{eq:explicitpaths}] which in turn depends on time $t$. In the quasi-adiabatic limit under a cyclic evolution in which \(\theta\) returns to its original value, the charge transport is purely geometrical and does not depend on the explicit time dependence of $\theta$.

In practice, with a number of points $N \sim 10$, one has 
a very accurate result for $\gamma_{C,S}$. In addition, although with a 
slower convergence with system size $L$, 
the exponential position operator can be used to arrive at the many-body polarization 
\cite{Resta1998,Aligia1999}, which is the one-point approximation of \cref{eq:berry}. The position operator defined in Ref. \onlinecite{Resta1998}
cannot be used for interacting systems with fractional filling
but can easily be extended \cite{Aligia1999}. The extension to 
$\gamma_S$ has been introduced in Ref. \onlinecite{Aligia2000}.
We use the following form of the charge- and spin-polarization:
\begin{align}
  P_C(\theta)=\frac{1}{2 \pi} \operatorname{Im} \ln \left\langle\Psi(\theta)\left|\hat{X}_C^{e}\right| \Psi(\theta)\right\rangle,\\
  \label{eq:manybodypol}
  P_S(\theta)=\frac{1}{2 \pi} \operatorname{Im} \ln \left\langle\Psi(\theta)\left|\hat{X}_S^{e}\right| \Psi(\theta)\right\rangle,
\end{align}
with \(\hat{X}_{C,S}^{e} = \exp \left[i (2 \pi / L) \Sigma_{j} j\left(\hat n_{j \uparrow} \pm \hat n_{j \downarrow}\right)\right] \) \cite{Aligia2000}.
\(2\pi P_{C,S}\) is equivalent to 
$\gamma_{C,S}$.

The thermodynamic phases of the IHM are distinguished by their values of the charge- and spin-Berry phases \(\gamma_C\) and \(\gamma_S\) \cite{Torio2001}.
More precisely, the Berry phases \((\gamma_C,\gamma_S)\) are quantized due to inversion symmetry and have the values \((\pi, \pi)_{\mathrm{MI}},(\pi, 0)_{\mathrm{SDI}}\) and \((0,0)_{\mathrm{BI}}\).
These quantized Berry phases arise in our models for \(\delta =0\). Additionally, due to a spin-rotation symmetry of $\pi$ around any axis perpendicular to the $z$-axis in spin space for all values of \(\delta\) and \(\Delta\), \(\hat H_\mathrm{IHZ}\) and \(\hat H_\mathrm{IH}\) can only have spin-Berry phases of \(0\) or \(\pi\), whereas \(\hat H_\mathrm{IHB}\) breaks this symmetry, allowing for arbitrary spin-Berry phases.

For all finite-system calculations, we use open-shell boundary conditions (periodic boundary conditions for 
a number of sites $L$ multiple of four, antiperiodic for even $L$ not a 
multiple of four) to allow for the resolution of gap closings.

\subsection{Real-time calculations}

For the finite-time calculation,
we parameterize the pump cycles with the time \(t\) as
\begin{align}
  \label[]{eq:finiteTcycle}
  \theta = \frac{2\pi}{T}t,
\end{align}
where \(T\) is the pump period.
The accumulated pumped charge [spin] at time $t$ is calculated via
\begin{align}
  \label{eq:pumpedcharge}
  Q_{[S]}(t)&=\int\limits_0^t d t'\langle\hat J_{[S]}(t')\rangle,
\end{align}
where the total particle and spin currents, averaged over two links are
\begin{align}
  \label{eq:current}
  \hat{J}&=\frac{i}{2} \sum_{j=1,2 ; \alpha = {\uparrow,\downarrow}}\left(J_{j} \hat{c}_{j,\alpha}^{\dagger} \hat{c}_{j+1,\alpha}-\text { h.c. }\right)\\
  \hat{J_S}&=\frac{i}{2} \sum_{j=1,2 ; \alpha = {\uparrow,\downarrow}}\left(\sigma_{\alpha,\alpha}^z J_{j} \hat{c}_{j,\alpha}^{\dagger} \hat{c}_{j+1,\alpha}-\text { h.c. }\right),
\end{align}
where $J_j=J(1+\delta(-1)^j)$.
In order to minimize transient non-adiabatic effects, the pumping is first started slowly via a quadratic ramp-up of the driving \cite{Privitera2018}.
We use a pumping period of \(TJ=50\), which is enough to ensure quasi-adiabaticity for the IHB and IHZ models.
For the IH model, strong finite-size effects \cite{Li2017} make Lanczos calculations unfeasible.
We therefore also use infinite-system density matrix renormalization group (DMRG) methods \cite{Schollwock2011,McCulloch2008} to calculate the pumped charge in time-dependent simulations. The ground state is calculated via the variational uniform matrix-product state (VUMPS) method \cite{Zauner-stauber2018}. The time-evolution is carried out via infinite time-evolving block decimation (iTEBD) \cite{Orus2008b}.
For the IH model, we use a period of \(TJ = 1000\) and a maximum bond dimension of \(\chi = 200\). All DMRG calculations are done using the ITensor library for Julia \cite{itensor}.

\section{Results}
\label{sec:results}

\subsection{Real-time calculations}
\begin{figure}[t]
  \centering
  \includegraphics{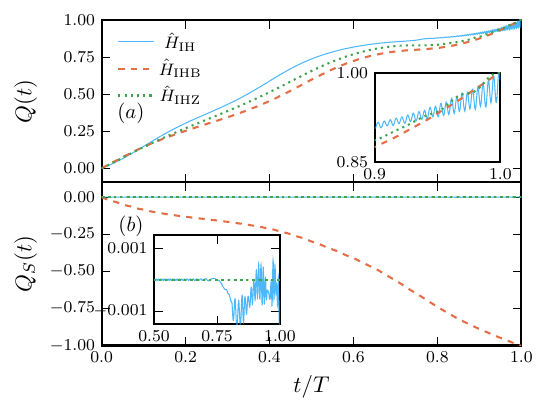}
  \mycaption{Real-time calculation of the pumped charge along \(\mathcal C_1\)}{(a): Both models \(\hat H_\mathrm{IHB}\) and \(\hat H_\mathrm{IHZ}\) show integer-quantized charge pumping. \(\hat H_\mathrm{IH}\) shows approximately quantized pumping. (b): \(\hat H_\mathrm{IHB}\) additionally shows quantized spin transport, whereas \(\hat H_\mathrm{IHZ}\) is a pure charge pump. Calculated via Lanczos for \(L=12\) and \(TJ=50\) for IHB and IHZ and for infinite-system size via iTEBD with \(\chi = 200\) and \(TJ = 1000\) for IH, with \(U/J=4\) for all models. The system parameters are: IH: \(R_\Delta/J = 1\), \(R_\delta=0.2\), \(\Delta_c/J=2.2\). IHB: \(R_\Delta/J = 2\), \(R_\delta=0.9\), \(\Delta_c/J=2.24\). IHZ: \(R_\Delta/J = 2\), \(R_\delta=0.9\), \(\Delta_c/J=2\).}
  \label{fig:finite}
\end{figure}

First, we consider a finite pumping period and demonstrate the quantized particle pumping in a time-dependent calculation of the integrated current for finite systems.
For \(\hat H_\mathrm{IHB}\) and \(\hat H_\mathrm{IHZ}\), the results for the pumped charge after one period along pump cycle \(\mathcal C_1\) from \cref{fig:cartoon_paths} are shown in \cref{fig:finite}(a) for a system size of \(L=12\) and \(TJ=50\). Both models show an integer-quantized pumping of a single charge. We have checked convergence with respect to the system size \(L\) for both models. In \cref{fig:finite} (b), the pumped spin of the same models and pumping path is shown. \(\hat H_\mathrm{IHZ}\) pumps no spin and is therefore a pure charge pump. \(\hat H_\mathrm{IHB}\) shows an integer-quantized pumping of a single spin along \(\mathcal C_1\).

\begin{figure}[t]
  \centering
  \includegraphics{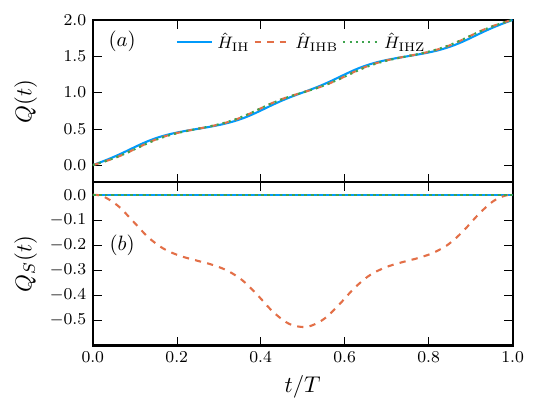}
  \mycaption{Lanczos calculation of the pumped charge along \(\mathcal C_2\) in a finite system.}{(a): All models show integer-quantized charge pumping of two particles per cycle. (b): \(\hat H_\mathrm{IHB}\) displays a finite spin current but a net-zero pumped spin. Calculated for \(L=12\), \(U/J=4\), \(TJ=50\), \(R_\Delta/J = 4\) and \(R_\delta=0.9\).}
  \label{fig:finiteRM}
\end{figure}

In contrast, for \(\hat H_\mathrm{IH}\) along \(\mathcal C_1\), finite-system calculations are not sufficient to overcome the large finite-size effects that arise from pumping through a gapless phase \cite{Li2017}.
We therefore employ infinite-system size calculations for this model. For a period of \(TJ = 1000\) and a maximum bond dimension of \(\chi = 200\), the results are presented in \cref{fig:finite} for \(\mathcal C_1\). We observe approximate integer charge pumping and no spin pumping. However, local spin oscillations arise when reaching the gapless point between the two critical points for \(t/T = 0.75\). This leads to an oscillatory behavior of the pumped charge. Interestingly, the envelope of these oscillations reaches a quantized value of one.
The calculations converge with increasing bond dimension until the gapless point. Beyond this, the local spin and charge oscillations show a strong dependence on both the bond dimension \(\chi\) and the pumping period \(T\). This suggests that in the thermodynamic limit, the quantized pumping may break down. This limit is only recovered for \(\chi\to \infty\) in infinite-system size DMRG.
Along \(\mathcal C_2\), all models exhibit quantized pumping of two particles and zero spin for \(TJ = 50\), which is shown for \(L=10\) in \cref{fig:finiteRM}.

We have checked the quantization for the first 20 pump cycles and find a very robust quantization for \(\mathcal C_2\) for all models as expected. In the infinite system, calculations for multiple pump cycles around \(\mathcal C_1\) indicate a significant deviation from quantization from the second pump cycle onward for \(\hat H_\mathrm{IH}\) and \(\hat H_\mathrm{IHZ}\). For a finite system, the pumped charge for both \(\hat H_\mathrm{IHB}\) and \(\hat H_\mathrm{IHZ}\) is integer-quantized for the first 20 pump cycles within experimental accuracies  [results not shown here].

\subsection{Energy gaps}
\label{sec:engaps}

\begin{figure}[t]
  \centering
  \includegraphics{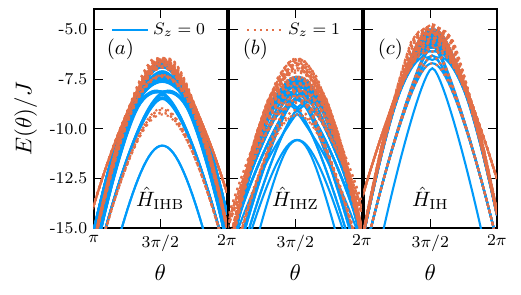}
  \mycaption{Energy levels along \(\mathcal C_1\).}{The lowest 50 energy levels along the path \(\mathcal C_1\) are plotted for the \(S_z=0\) and \(S_z = 1\) subspaces. \(\hat H_\mathrm{IHB}\) has a non-degenerate ground state that is separated from the rest of the spectrum. \(\hat H_\mathrm{IHZ}\) has a twofold-degenerate ground state at \(\theta = 3\pi/2\) that is separated from the rest of the spectrum. \(\hat H_\mathrm{IH}\) becomes fully gapless for \(L\to\infty\). Calculated for \(L=12\), \(U/J=4\), \(R_\Delta/J = 2\), \(R_\delta=0.9\) and \(\Delta_c/J=2.24\).}
  \label{fig:tower}
\end{figure}

In order to understand the real-time simulations, we consider instantaneous measures for all models. We calculate the lowest 50 eigenenergies for all three models in the symmetry sectors \(S_z = 0\) and \(S_z=1\) for \(L=12\) along the path \(\mathcal C_1\). The results are shown in \cref{fig:tower} for \(\theta \in [\pi,2\pi]\) which corresponds to half a pump cycle  \(\mathcal C_1\) of \cref{fig:cartoon_paths}. At \(\theta=3\pi /2\), the path is in between the two critical points along the \(\delta = 0\) line.

The ground state of \(\hat H_\mathrm{IHB}\) is non-degenerate for all values of \(\theta\) and separated from the rest of the spectrum by a robust gap. The ground state of \(\hat H_\mathrm{IHZ}\) is twofold-degenerate for \(L\to \infty\) in the MI phase, which extends to a region around
$\Delta=\delta=0$. 
Proceeding as in App.\;\ref{sec:app} the ground state energy becomes
\begin{equation}
E_{MI} = - J_z/4  - 4 \tilde U J^2 (1+\delta^2) / (\tilde U^2-4 \Delta^2),
\label{emi}
\end{equation} for both Néel-like states, where \(\tilde U = U + 3 J_z/4\). However, the two crossing levels that make up this ground-state manifold are separated from the rest of the spectrum. The real-time simulations indicate that this is sufficient to ensure quantized particle transport for at least the first few pump cycles.
In contrast, \(\hat H_\mathrm{IH}\), which becomes the regular ionic Hubbard model at \(\delta =0\), becomes fully gapless in the thermodynamic limit, since the spin gap vanishes in the MI at \(\delta=0\) \cite{Fabrizio1999}. We therefore expect that in the thermodynamic limit, the pumping will ultimately break down in the IH case, consistent with \cite{Nakagawa2018}. In this sense, we believe that for a finite system, which is relevant for ultra-cold atomic gas experiments, the finite-size gap can be used to protect the pumping of an integer amount of particles for a few pump cycles.

\begin{figure}[t]
  \centering
  \includegraphics{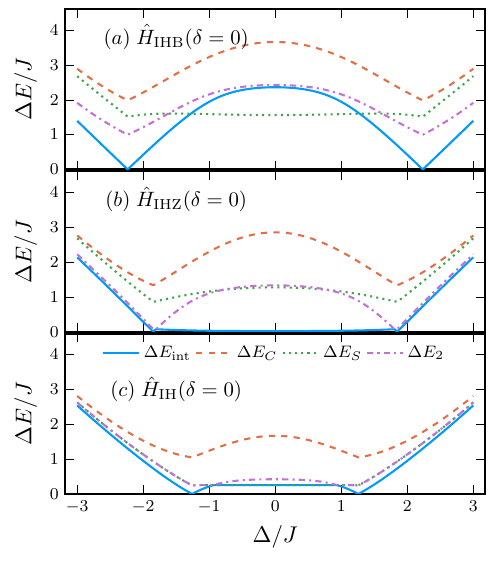}
  \mycaption{Energy gaps along the \(\delta =0\) line.}{(a): In the center between the two critical points, \(\hat H_\mathrm{IHB}\) is fully gapped with the smallest gap being the spin gap. (b): For \(\hat H_\mathrm{IHZ}\), the second internal gap approaches the finite spin gap. The internal gap vanishes due to the twofold-degenerate ground state. (c): For \(\hat H_\mathrm{IH}\) the spin gap is equivalent to the internal gap, which becomes zero in the thermodynamic limit. The charge gap is finite for all models. Calculated for \(L=12\) and \(U/J=4\).}
  \label{fig:engaps}
\end{figure}

In \cref{fig:engaps}, the energy gaps defined in \cref{sec:model} are shown along the \(\delta =0\) line, where we expect a gapless phase between the two critical points of the IHM. The gaps are calculated for various system sizes and shown for \(L=12\). For all three models, the charge gap is finite for all system sizes, which is a necessary condition for quantized charge transport. For \(\hat H_\mathrm{IHB}\), all gaps are finite except at the critical points. The spin gap becomes the smallest gap between the critical points and is on the order of the staggered magnetic field term,  which is independent of the system size. Therefore, the topologically protected charge pumping in this model is robust for all system sizes.

The internal gap of \(\hat H_\mathrm{IHZ}\) vanishes between the two critical points, as was already observed in \cref{fig:tower}. This is due to a twofold-degenerate ground-state manifold between a N\'eel-like state 
($\uparrow$ $\downarrow$ $\uparrow$ $\downarrow$ …) and an anti-N\'eel like one
($\downarrow$ $\uparrow$ $\downarrow$ $\uparrow$ …) in the thermodynamic limit, which is unaffected by the Ising term. The second internal gap stays finite, which shows that the ground-state manifold is separated by a gap from the rest of the spectrum.

For \(\hat H_\mathrm{IH}\), the internal gap is of the order of a finite-size gap and converges very slowly with system size. The internal gap vanishes at the MI to SDI transition, but remains finite in the SDI phase, which has been shown in \cite{Manmana2004}. The SDI to MI transition is characterized by a crossing of excited energy levels. The excited even singlet crosses with the excited odd triplet, which has less energy in the MI phase \cite{Torio2001}. Specifically, the internal gap becomes the spin gap in the MI phase.
This is  due to a crossing of energy levels with opposite inversion symmetry. An odd singlet is the ground state in the SDI and MI phases, 
while in the BI, the ground state is an even singlet \cite{Torio2001}. According to conformal field theory, the spin gap scales as 
$ 2 \pi v_s/L$, where $v_s$ is the spin velocity \cite{Nomura1994c}. Therefore, the IHM becomes gapless in the thermodynamic limit.

\subsection{Berry phase and many-body polarization}

\begin{figure}[t]
  \centering
  \hspace{-5mm}
  \includegraphics[width = 1.05\columnwidth]{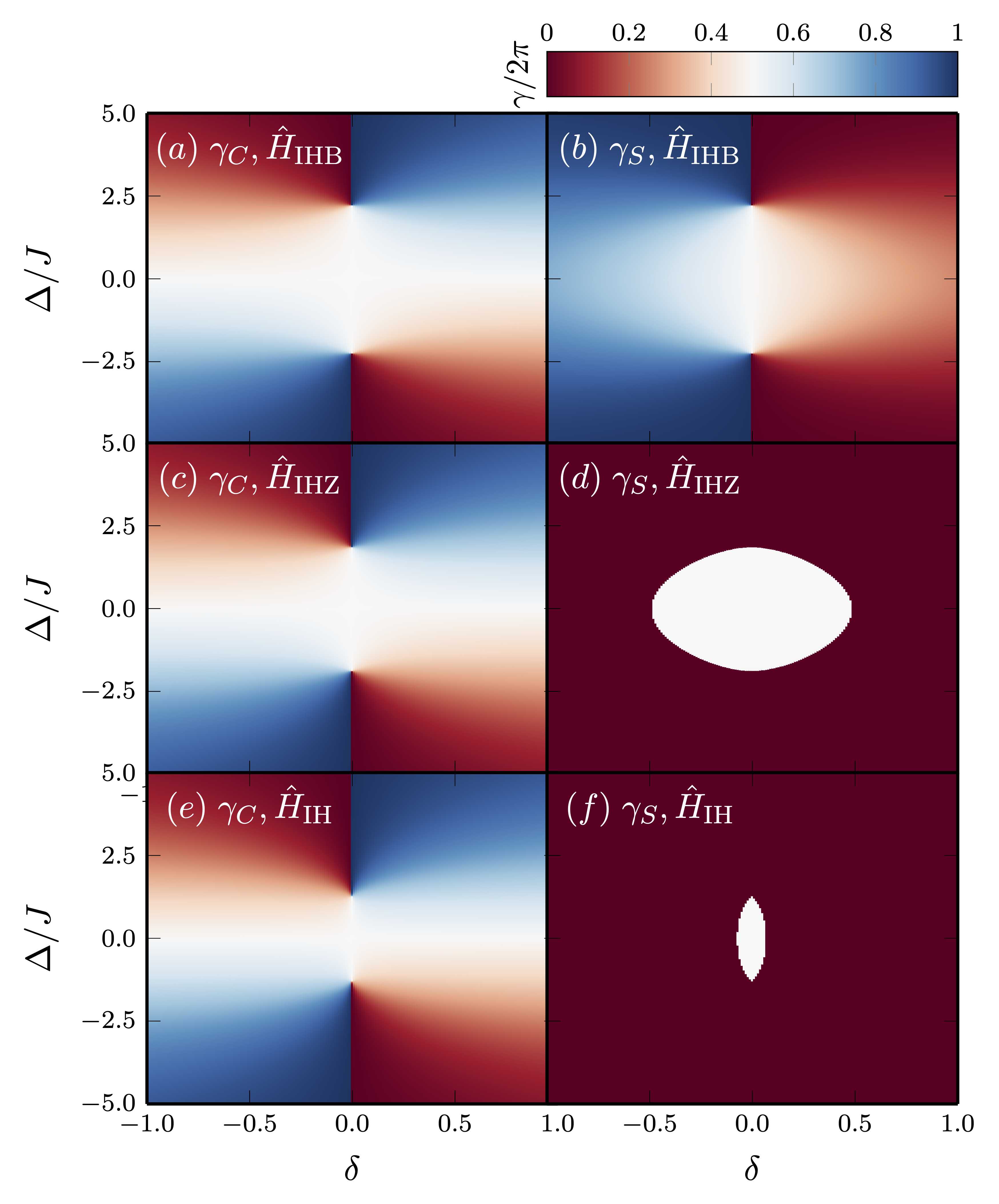}
  \mycaption{Charge- and spin-Berry phases in the \(\delta-\Delta\) plane.}{(a,c,e): Charge-Berry phase \(\gamma_C\). (b,d,f): Spin-Berry phase \(\gamma_S\). (a,b): \(\hat H_\mathrm{IHB}\), (c,d): \(\hat H_\mathrm{IHZ}\), (e,f): \(\hat H_\mathrm{IH}\). Calculated for \(L=10\) and \(U/J=4\).}
  \label{fig:berry2}
\end{figure}

We now address the topology of the three models. In particular, we are interested in the charge- and spin-Berry phases, which give information on the pumped charges and spins in a quasi-adiabatically driven system. We use the many-body polarization in \cref{eq:manybodypol} to calculate the Berry phases for a finite system of \(L=10\).
The results for the charge- [spin-] Berry phases are shown in \cref{fig:berry2} (a,c,e) [(b,d,f)]. For all models, the charge-Berry phases are well-defined and smooth everywhere except for the two critical points. Notice that the branch cuts that emerge from the critical points are \(\Delta\gamma_{C,S} = \pm2\pi\) and therefore well-defined. The position of the branch cuts can be changed via a gauge transformation. Physically relevant information is only encoded in the total Berry phase picked up along a closed path.

\(\hat H_\mathrm{IHB}\) has a well-defined and smooth spin-Berry phase. Notice that around the upper singularity, the sign of the spin-Berry phase is opposite to the charge-Berry phase. This means that encircling one critical point
pumps both spin and charge.  More specifically, pumping around the upper [lower] critical point pumps only a single spin down [up] particle.

The spin-Berry phase for \(\hat H_\mathrm{IHZ}\) and \(\hat H_\mathrm{IH}\) only has the values \(\gamma_S \in \{0,\pi\}\). The value of \(\pi\) is realized between the critical points for both models as is expected for the IHM. For \(\gamma_S=\pi\), the ground state is in the MI phase. The quantization arises due to the spin-rotation symmetry in these two models which maps \(\gamma_S \mapsto -\gamma_S\). For \(\hat H_\mathrm{IH}\), the spin-Berry phase is expected to be nonzero only for \(\delta \neq 0\) in the thermodynamic limit, since a finite \(\delta\)
breaks the inversion symmetry and leads to a finite BOW order parameter (see App. \ref{sec:app}). The small lentil shape as seen in \cref{fig:berry2} (f) is therefore likely a finite-size effect. 
For \(\hat H_\mathrm{IHZ}\), the transition between dimerized phase and Mott phase happens at finite \(\delta\). A similar transition has recently been observed in dimerized XXZ Hamiltonians \cite{Tzeng2016a}. The value of \(\delta\) where the transition happens decreases with increasing system size for small systems (not shown here). Perturbation theory along the lines of App.\;\ref{sec:app}  [see Eq. (\ref{emi})] indicates that this region remains 
finite, though. In real-time simulations, the jump in \(\gamma_S\) has no effect on the quantization of pumped charge for the IHZ model. We therefore argue that quantized particle pumping without spin pumping is possible around a single critical point in this model.

\begin{figure}[]
  \centering
  \includegraphics{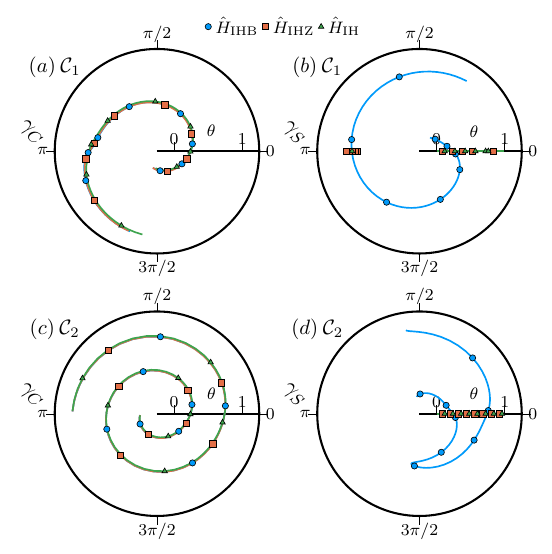}
  \mycaption{Charge- and spin-Berry phases along the pump cycles \(\mathcal C_1\) and \(\mathcal C_2\).}{(a,c): Charge-Berry phase \(\gamma_C\). (b,d): Spin-Berry phase \(\gamma_S\). Calculated for \(L=10\) and \(U/J=4\). $\mathcal C_1$: \(R_\Delta/J = 2\), \(R_\delta=0.9\), \(\Delta_c/J=2.24\). $\mathcal C_2$: \(R_\Delta/J = 4\), \(R_\delta=0.9\).}
  \label{fig:polpol1}
\end{figure}
Figure ~\ref{fig:polpol1} shows the charge- [spin-] Berry phases \(\gamma_C\) [\(\gamma_S\)] for both paths \(\mathcal C_1\) and \(\mathcal C_2\) for all three models as the angular variable in a polar plot \cite{Hayward2021a}. This is done because the winding of the Berry phase is equal to the pumped charge. The charge-Berry phase shows an integer winding for both \(\mathcal C_1\) and \(\mathcal C_2\) for all models, which mirrors the results from \cref{fig:berry2}. The spin-Berry phase only shows a well-defined winding of one in the case of \(\mathcal C_1\) and zero for \(\mathcal C_2\) for \(\hat H_\mathrm{IHB}\). This is consistent with an interacting Rice-Mele pump that pumps one charge per species and no spins. For \(\mathcal C_2\), the IHZ and as IH models show a smooth spin-Berry phase with no winding as well, as long as the lentil shape of \(\gamma_S=\pi\) is surrounded completely.
The \(\mathcal C_1\) paths inevitably go through the spin-transition and therefore show a discontinuity in the spin-Berry phase. This means that the spin-Berry phase no longer has a well-defined winding and no adiabatic spin-transport should be possible in an infinite-system. However, in practice, we see that in real-time simulations, \(\hat H_\mathrm{IHZ}\) effectively behaves as if \(\gamma_S\) has a well-defined zero winding, at least for the first pump cycle. This is true for both finite-size and infinite-system calculations (the latter not shown here). We therefore expect this model to be well-behaved in ultra-cold atom experiments with finite particle numbers.


\section{Summary and discussion}
\label{sec:summary}

We showed that it is possible to split up the degeneracy of a two-component Rice-Mele model via a Hubbard interaction term. We presented three concrete models to achieve this that are based on an interacting two-component Rice-Mele model with a shifted pump cycle: A spin-SU(2) symmetric model, which realizes the ionic Hubbard model during the pump cycle (IH), a model with an additional staggered magnetic field (IHB), and a model with an additional Ising term (IHZ).
We confirmed the quantization of the pumped charge via finite and infinite-system real-time calculations and instantaneous measures for periodic boundary conditions.

The quantization is most robust in the IHB case, which is robustly gapped everywhere. As a consequence, both charge- and spin-Berry phases are well-defined everywhere except at the critical points. However, the staggered field leads to an additional non-zero quantized spin-pumping.

For IHZ, quantization holds for the first couple of pump cycles for experimentally relevant time scales in a finite system, despite the twofold-degenerate ground-state. While the charge-Berry phase is well-defined as in the IHB case, the spin-Berry phase jumps at a finite value of the hopping modulation.

The ionic Hubbard model, which is visited during pump cycles in the IH case, features a Mott phase with a vanishing spin gap to a continuum of excitations that we pump through, which should lead to an eventual breakdown of quantized particle transport. However, a clear remnant of the underlying topology is preserved and the pumped charge is quantized approximately in the first cycle. This is consistent with the well-defined charge-Berry phase in this case. The spin-Berry phase shows a jump similar to the IHZ case, which is only expected at zero hopping modulation in the thermodynamic limit. Unlike the IHB, the spin current is manifestly zero for the IH and IHZ.

In \cite{Nakagawa2018}, Nakagawa \textit{et al.} theoretically study the same model and interpret their results in terms of a breakdown of quantized particle pumping due to the repulsive Hubbard interaction. 
For open boundary conditions, the many-body polarization \cite{King-smith1993,Resta1994,Ortiz1996,Resta1998} shows a quantized
jump due to the emergence of edge states in an OBC system \cite{Hatsugai2016}. For finite interaction strength, these edge-state contributions are shown to split up along the pump cycle which eventually leads to a breakdown of quantized pumping.
In our context of splitting degenerate points, the breakdown in the interacting two-component Rice-Mele model is seen when the splitting of the single degeneracy at \(\Delta = 0\) into two critical points at \(\pm \Delta_c\) due to the Hubbard interaction surpasses the \(\Delta\)-radius of the origin-centered pumping path in the \(\delta-\Delta\) plane. Therefore, the pumping path \(\tilde{\mathcal{C}}_2\) chosen by Nakagawa \textit{et al.} indeed encounters a gapless phase between the two spin-critical points \(\pm\Delta_S\) twice but most importantly, does not encircle an isolated singularity and hence no charge is pumped. We argue that this primarily constitutes a transition from pumping a quantized number of two to zero particles during initial pump cycles, while the breakdown due to the spin-gapless line will manifest itself after sufficiently many pump cycles.

We believe that the same mechanism of this interaction-induced splitting of the degeneracies while keeping the pump cycle fixed is at the heart of the results reported in the recent experimental work by Walter \textit{et al.} \cite{Walter2022} as well.
Of course, the SU\((2)\) symmetric model possesses a gapless line, which in principle should prevent quantized pumping altogether. Our numerical results, however, show that this source of a breakdown is very unlikely to manifest itself on initial pump cycles or finite systems even for a uniform system. In a system with an open charge gap but a vanishing spin gap somewhere along the pump cycle, one first expects spin excitations. A heating up of the charge sector may not immediately occur. How exactly the breakdown of quantized pumping due to gapless spin excitations behaves as a function of system size and which time scales are relevant is an open question and demands further research. With regards to the interpretation of the experiment by Walter \textit{et al.}, one should also stress that their system confines particles in a harmonic trap and, as a consequence, arbitrarily slow pumping will not lead to quantized pumping anyway, because the metallic edges will hybridize and hence be coupled by a finite tunneling rate \bibnote[Masterarbeit]{E. Bertok,  Master thesis, Georg-August Universit\"at G\"ottingen (2019)}. 

We would further like to emphasize that the realization of a Mott insulator per se does not preclude the possibility of quantized pumping, which is supported by our results for the IHB and IHZ models and the results for pumping in a bosonic MI \cite{Hayward2018}.
Furthermore, it should be noted that based on our results for the IH model, which is most easily realized experimentally, quantized transport around a single critical point may require considerably slower pumping than is currently possible in ultra-cold atom experiments.

Interesting results are expected when pumping through the SDI phase directly.  For example, Nakagawa \textit{et al.} report on the possibility of fractional pumping in this case \cite{Nakagawa2018}.
In the present work, we do not see any effect on the pumping when going through the SDI phase. However, we have not further pursued this question due to the problem of pumping close to the degeneracies and consequently large inherent finite-size effects.

\section*{Acknowledgments}
We are grateful to Jan Albrecht and Michael Fleischhauer for fruitful discussions
and we thank Masaya Nakagawa for comments on a previous version of the manuscript.
This research was funded by the Deutsche Forschungsgemeinschaft
(DFG, German Research Foundation) via Research Unit FOR 2414
under project number 277974659 
and by PICT 2017-2726 and PICT 2018-01546 of the ANPCyT, Argentina. 
A.A.A. thanks the Alexander von Humboldt Foundation for support.






\appendix
\section{Spin gap, polarization and bond order parameter of the IHM for small hopping}
\label{sec:app}
We write the Hamiltonian in the form

\begin{eqnarray}
\hat H_{IH} &=&\hat H_{0}+\hat H_{J},  \nonumber \\
\hat H_{0} &=&\sum\limits_{i}\left[ \Delta (-1)^{i}\hat n_{i}+U\hat n_{i\uparrow
}\hat n_{i\downarrow }\right] ,  \nonumber \\
\hat H_{J} &=-J&\sum\limits_{i\sigma } \left[1+(-1)^{i}\delta \right] \left(
\hat c_{i+1\sigma }^{\dagger }\hat c_{i\sigma }+\text{H.c.}\right) ,  \label{hp}
\end{eqnarray}%
where $\hat n_{i\sigma }=\hat c_{i\sigma }^{\dagger }\hat c_{i\sigma }$ and $%
\hat n_{i}=\hat n_{i\uparrow }+\hat n_{i\downarrow }$.
The calculations below correspond to
the static situation.

We describe the calculation of the polarization $P_C$  and the order parameter of the BOW phase to
lowest non-trivial order in $\hat H_{J}$ for a ring of $L$ sites, starting 
from either the BI or MI phases. The SDI phase is out of the reach of the validity of the present perturbative treatment.

We perform a canonical transformation similar to the one that 
transforms the Hubbard model at half filling to a Heisenberg model.
To second order in $\hat H_{J}$, the transformed Hamiltonian is \cite{Aligia2004}

\begin{eqnarray}
\hat{\tilde{H}}&=&\hat Pe^{-\hat S}\hat He^{\hat S}\hat P \nonumber \\
&=&P\left\{ \hat H+\left[ \hat H,\hat S\right] +\frac{1}{2}\left[ \left[ \hat H,\hat S\right] ,\hat S\right] +...\right\} \hat P \nonumber \\
&\simeq& \hat P\left\{ \hat H_{0}+\left[
\hat H_{J},\hat S\right] \right\} \hat P,  \label{ht}
\end{eqnarray}
where $\hat P$ is the projector over the ground state $|g_{0}\rangle $ of $\hat H_0$ and in the last equality we have used

\begin{equation}
\hat H_{J}+\left[ \hat H_{0},\hat S\right] =0,  \label{s1}
\end{equation}
to eliminate terms linear in $\hat H_{J}$ in $\hat{\tilde H}$. Using this equation,
the matrix elements of $\hat S$ between eigenstates of $\hat H_{0}$\ are easily
determined:

\begin{equation}
\langle n|\hat S|m\rangle =\frac{\langle n|\hat H_{J}|m\rangle }{E_{n}-E_{m}}.
\label{sdef}
\end{equation}%
Note that $\hat S$ is anti-hermitian ($\langle m|\hat S|n\rangle ^{\ast }=-\langle
n|\hat S|m\rangle $).

Starting from the BI phase, $\hat{\tilde H}$ is trivial and reduces to the projector on the non-degenerate ground state $|g_{0}\rangle $. Instead,
starting from the MI phase, $|g_{0}\rangle $ is degenerate and 
$\hat{\tilde H}$ takes the form of a
Heisenberg chain with
alternating exchange parameters $J_{1(2)}=4J(t \pm \delta
)^{2}U/(U^{2}-4\Delta ^{2})$. This effective model can be written in the
form \cite{Aligia2004,Nakagawa2018}

\begin{equation}
\hat H_{\text{Heis}}=\sum\limits_{i}\left[ J_{\text{Heis}}+(-1)^{i}x\right] 
\mathbf{\hat S}_{i+1}\cdot \mathbf{\hat S}_{i}.  
\label{heis}
\end{equation}
with $J_{\text{Heis}}=(J_1+J_2)/2$ and 
$x=4J^2\delta\;U/(U^{2}-4\Delta ^{2})$. 
Using previous results on this model using bosonization \cite{Cross1979},
one knows that a gap proportional to $\delta^{2/3}$ opens for small
$\delta \neq 0$.

The expectation values of the occupancies can be calculated in the new basis as

\begin{equation}
\left\langle \hat n_{i}\right\rangle =\left\langle g\left\vert\hat n_{i}\right\vert g\right\rangle
=\left\langle g\right\vert e^{\hat S}e^{-\hat S}\hat n_{i}e^{\hat S}e^{-\hat S}\left\vert g\right\rangle =\left\langle
g_{0}\right\vert\hat{\tilde{n}}_{i}\left\vert g_{0}\right\rangle ,  \label{ni0}
\end{equation}
where $|g \rangle = e^{\hat S}|g_{0} \rangle$ and 
to second order in $\hat H_{J}$

\begin{equation}
\hat{\tilde{n}}_{i}=\hat Pe^{-\hat S}\hat n_{i}e^{\hat S}P\simeq \hat P\left\{ \hat n_{i}+\left[ \hat n_{i},\hat S\right] +%
\frac{1}{2}\left[ \left[ \hat n_{i},\hat S\right] ,\hat S\right] \right\} \hat P  \label{ntrans}
\end{equation}

Since $\left\langle g_{0}\right\vert\hat S\left\vert g_{0}\right\rangle =0$ and $\hat n_{i}|g_{0}\rangle
=\left\langle \hat n_{i}^{0}\right\rangle |g_{0}\rangle $, where $\left\langle
\hat n_{i}^{0}\right\rangle =\left\langle g_{0}|\hat n_{i}|g_{0}\right\rangle$
, the second term between brackets does not contribute and then 
\begin{equation}
\left\langle \hat n_{i}\right\rangle =\left\langle \hat n_{i}^{0}\right\rangle
+\left\langle g_{0}\left\vert\left\langle \hat n_{i}^{0}\right\rangle
\hat S^{2}-\hat S\hat n_{i}\hat S\right\vert g_{0}\right\rangle .  \label{ni2}
\end{equation}

Taking matrix elements of the second term, it is clear that only excited
states for which $n_{i}\neq \left\langle \hat n_{i}^{0}\right\rangle $ contribute to it.

\subsection{Polarization in the band insulating phase.}

\label{bi}

In the BI phase, $\left\langle \hat n_{i}^{0}\right\rangle=0$ or 2 and all
intermediate states have $n_{i}=1$. Then, \cref{ni2} leads to

\begin{equation}
\left\langle \hat n_{i}\right\rangle =\left\langle \hat n_{i}^{0}\right\rangle +\left(
1-\left\langle \hat n_{i}^{0}\right\rangle \right) \Sigma _{k}^{\prime}\frac{%
\left\vert \langle k|\hat H_{J}|g_{0}\rangle \right\vert ^{2}}{\left(
E_{k}-E_{0}\right) ^{2}},  \label{ni3}
\end{equation}
where the sum is restricted to the two excited states $|k\rangle$
of $\hat H_{0}$ obtained after applying $\hat H_{J}$ to $|g_{0}\rangle $ 
for which $n_{i}=1.$
One realizes that for positive $\Delta$ there are $L/2$ hops to the
right for each spin with matrix element $-J(1-\delta)$ and $L/2$ hops to the left for each spin with matrix element $-J(1+\delta)$. For negative 
$\Delta$, the situation is the opposite. Therefore, the change 
in polarization with respect to the BI phase for $\hat H_J=0$ is

\begin{equation}
\Delta P_C^{\text{BI}}=-\frac{4J\delta\:\text{sgn}(\Delta)}{\left( 2\Delta -U\right) ^{2}}.
\label{dpbip}
\end{equation}

To compare with numerical calculations of the charge Berry 
phase,
we have chosen $U=0$, $\Delta =10,$ $J=1,$ $\delta
=0.1$. Eq. (\ref{dpbip}) gives $\Delta P_{BI}^{\text{PBC}}=-10^{-3}$. The
numerical calculation for $L=6,$ 8 and 10 gives $\Delta P_{\text{PBC}%
}=(-9.83\pm 0.01)\times 10^{-4}$. Both results differ by less than 2\%.

\subsection{Polarization near the Mott insulating phase.}

\label{mi}

The calculation of the polarization in this case is more difficult due to
the spin structure of $|g_{0}\rangle $. In particular, there can't be any
nearest-neighbor hopping if the spins of the electrons of the sites involved are parallel. Therefore the result depends on spin correlation functions.
From Eq. (\ref{ni2}), one realizes that the contribution of the state in which an
electron is displaced from site 1 to 2 is proportional to the probability
that the spins of sites 1 and 2 form a singlet, since this hopping is not
possible for triplets. The projector on the singlet state is 
$\left( 1/4-\mathbf{\hat S}_{1}\cdot \mathbf{\hat S}_{2}\right) $, and the matrix element for the
singlet has a factor $\sqrt{2}$. Explicitly:

\begin{eqnarray}
&&-J(1-\delta )\sum\limits_{\sigma }\hat c_{2\sigma }^{\dagger }\hat c_{1\sigma }\left[ 
\frac{1}{\sqrt{2}}\left( \hat c_{1\uparrow }^{\dagger }\hat c_{2\downarrow }^{\dagger
}-\hat c_{1\downarrow }^{\dagger }\hat c_{2\uparrow }^{\dagger }\right) |0\rangle %
\right] \nonumber \\
&&=-\sqrt{2}J(1-\delta )\hat c_{2\uparrow }^{\dagger }\hat c_{2\downarrow
}^{\dagger }|0\rangle .  \label{ms}
\end{eqnarray}
Thus, proceeding as before, the change in polarization with respect to the MI phase with  $J=0$ (with all $\left\langle n_{i}\right\rangle =1$) is

\begin{eqnarray}
\Delta P_C &=& \frac{8U\Delta J^2}{(U^{2}-4\Delta ^{2})^{2}} \{ (1+\delta
)^{2}\left( 1/4-\left\langle \mathbf{\hat S}_{3}\cdot \mathbf{\hat S}_{2}\right\rangle
\right) \nonumber \\
&&-(1-\delta )^{2}\left( 1/4 -\left\langle \mathbf{\hat S}_{1}
\cdot \mathbf{\hat S}_{2}\right\rangle \right) \}.  \label{dpmi}
\end{eqnarray}
For $\delta \rightarrow 0$, the correlation functions for all links are the
same and 
$1/4-\left\langle \mathbf{\hat S}_{i}
\cdot \mathbf{\hat S}_{i+1}\right\rangle=\ln 2$ in the thermodynamic limit \cite{DesCloizeaux1962}. Therefore, in this limit 
$\delta \rightarrow 0$, $L\rightarrow \infty $,  the above expression can be simplified to

\begin{eqnarray}
\Delta P_C &\simeq &\frac{8U\Delta J^2}{(U^{2}-4\Delta ^{2})^{2}}\left( 4\delta
\ln 2+O_{B}\right) ,  \nonumber \\
O_{B} &=&\left\langle (\mathbf{\hat S}_{1}-\mathbf{\hat S}_{3})\cdot \mathbf{\hat S}_{2}\right\rangle   \label{dpd0}
\end{eqnarray}%
$O_{B}$ is the dimer order parameter of $\hat H_{\text{Heis}}$
Using the Hellman-Feynman theorem 
$\partial E/\partial x=(L/2)O_{B}$, where  
$E=\left\langle \hat H_{\text{Heis}}\right\rangle $. From bosonization \cite{Cross1979} and numerical \cite{Okamoto1986}
results one knows that $E\sim x^{4/3}$ and then 
$O_{B}\sim \delta ^{1/3}$. 
Therefore, the change in the dimer order parameter with $\delta $ dominates $\Delta P$ for very small $\delta $. 

For a comparison with numerical calculations we take $J=1$, $U=10$, 
$\Delta=\delta =0.1$. This leads to $J_{2}/J_{1}=0.67$ for which 
$O_{B}\simeq 0.517$
according to Fig. 1 of Ref. \cite{Paul2017}.  Approximating $1/4-\left\langle 
\mathbf{\hat S}_{1(3)}\cdot \mathbf{\hat S}_{2}\right\rangle =\ln 2\pm O_{B}/2$, 
Eq.\;(\ref{dpd0})  gives $\Delta P=5.24\times 10^{-4}$. From the numerical
calculation of the Berry phases we obtain $\Delta P_C=z\times 10^{-4}$, with 
$z=4.73$, 5.17 and 5.38 for $L=6$, 8 and 10 respectively in reasonable
agreement with the above estimation.

\subsection{Bond-order parameter}

The parameter of the bond-order wave (BOW) can be defined as

\begin{eqnarray}
O_{\text{BOW}} &=&\frac{1}{L}\sum_i(-1)^{i}\left\langle \hat b_{i}\right\rangle , 
\nonumber \\
\hat b_{i} &=&\hat c_{i+1\sigma }^{\dagger }\hat c_{i\sigma }+\text{H.c.},  
\label{obow}
\end{eqnarray}
where $\left\langle \hat b_{i}\right\rangle $ is the expectation value of the
hopping between sites $i$ and $i+1$. For $\delta =0$, odd and even bonds are
equivalent and therefore $O_{\text{BOW}}=0$ for any finite system with an
even number of bonds. In the thermodynamic limit within the SDI\ phase there
is precisely a spontaneous symmetry breaking and the system "chooses" one of
two possible degenerate states with opposite $O_{\text{BOW}}$ \cite{Fabrizio1999}. The SDI phase is out of the reach of validity of the present perturbative
treatment. For $\delta \neq 0$, $O_{\text{BOW}}\neq 0$ and therefore the bond-order parameter can be analyzed by
perturbation theory in $H_J$ starting from the BI and MI phases. Using Eq.
(\ref{ntrans}) for $\hat O=\hat b_{i}$, it is easy to see that for both phases the
first non-trivial contribution is the linear one in $\hat S$ (the operator that
acts first between $\hat b_{i}$ and $\hat S$ leads so some excited state and the other
returns to the ground state). Thus

\begin{equation}
\hat{\tilde{b}}_{i}\simeq \hat P\left[ \hat b_{i},\hat S\right] \hat P  \label{btrans}
\end{equation}%
Doing the calculation for the non-degenerate ground state of the BI
phase using Eq. (\ref{sdef})  one obtains

\begin{equation}
\left\langle \hat{\tilde{b}}_{i}\right\rangle _{\text{BI}}=4J \frac{1+(-1)^{i}\delta 
}{2\Delta -U}.  \label{birbi}
\end{equation}%
Inserting this expression into Eq. (\ref{obow}) we obtain 

\begin{equation}
O_{\text{BOW}}^{\text{BI}}=\frac{4\delta J}{2\Delta -U}.  \label{bowbi}
\end{equation}

In the other phases, the calculation is more complicated because of the
structure of the ground state in which spin flips are possible. Proceeding in
a similar way as in Ref. \cite{Aligia2004} and above, we obtain 

\begin{equation}
\left\langle \hat{\tilde{b}}_{i}\right\rangle=\left( 1-4\left\langle 
\mathbf{\hat S}_{i}\cdot \mathbf{\hat S}_{i+1}\right\rangle \right) UJ
\frac{1+(-1)^{i}\delta }{U^{2}-4\Delta ^{2}}.  \label{birmi}
\end{equation}
Inserting this in Eq. (\ref{obow}) one obtains

\begin{equation}
O_{\text{BOW}}=\frac{2JUO_{B}}{U^{2}-4\Delta ^{2}}+\frac{UJ\delta 
}{U^{2}-4\Delta ^{2}}(1-2\left\langle (\mathbf{\hat S}_{3}+\mathbf{\hat S}_{1})\cdot 
\mathbf{\hat S}_{2}\right\rangle ),  \label{bowmi}
\end{equation}
where $O_{B}$ was discussed above. For $\delta \rightarrow 0$, $O_{B}\sim
\delta ^{1/3}$ and therefore, the first term is the leading one. This means
that in this limit the bond-order parameter is proportional to the dimer
order parameter of the Heisenberg chain with alternating exchange parameters.


\bibliographystyle{new-apsrev}
%

\end{document}